# Magnetoresistance and percolation in the LaNi$_{1-x}$Co$_x$O$_3$ solid solution

J. Androulakis,[a,*] Z. Viskadourakis,[a] N. Katsarakis,[a] and J. Giapintzakis [a,c,*]

[a]Institute of Electronic Structure and Laser, Foundation for Research and Technology – Hellas, P.O. Box 1527, Vasilika Vouton, 711 10 Heraklion, Crete, Greece

[c]Department of Materials Science and Technology, University of Crete,

P.O. Box 2208, 710 03, Heraklion, Crete, Greece

**Abstract**

A detailed study of the zero-field electrical resistivity and magnetoresistance for the metallic members of the LaNi$_{1-x}$Co$_x$O$_3$ solid solution with 0.3≤$x$≤0.6 is reported. The low temperature resistivity of the compounds with 0.3≤$x$≤0.5 exhibits a logarithmic dependence that is characteristic of systems with spin fluctuations. It is suggested that the effect of the magnetic field dependence on the spin fluctuations plays a vital role in determining the magnetoresistive behavior of these compounds. Concrete experimental evidence that classify the chemically induced metal-to-insulator transition ($x_c$=0.65) as a percolative phenomenon is provided. The resistivity data for the $x$=0.6 metallic compound are analyzed in the framework of cluster percolation threshold theory. The results of this analysis are consistent with the suggestion that the growth of magnetic metallic clusters in the presence of a magnetic field is mainly responsible for the observed giant magnetoresistance effect at low temperatures for the compounds with $x$≥0.6.

*Corresponding authors: J.A.: **amado@iesl.forth.gr**, J.G.: **giapintz@iesl.forth.gr**

# I. Introduction

Large negative magnetoresistance (MR), a technologically important physical property, has been reported for a variety of systems, such as metallic multilayers, granular intermetallic alloys and perovskite-type oxides (e.g. manganites) [1]. Several members of the LaNi$_{1-x}$Co$_x$O$_3$ solid solution also belong to the exceptional class of materials that exhibit giant magnetoresistance (GMR), however, no consistent explanation for this effect has been proposed to date [2].

The end members of this solid solution series are LaNiO$_3$ and LaCoO$_3$. LaNiO$_3$ is a strongly correlated metallic Stoner enhanced paramagnet down to 4.2 K [3] and LaCoO$_3$ is an insulator that exhibits interesting magnetic properties such as a Co-based thermally induced spin state transition at low temperatures [4]. Co doping of LaNiO$_3$ introduces disorder since the d-orbitals of Co and Ni ions differ by ~1 eV in energy and results in a metal-to-insulator transition at a critical composition of $x_c$=0.65 [5]. Members of the series with $x \geq 0.3$ exhibit enhanced ferromagnetic interactions below 60 K [6], which we have attributed to a cluster glass behavior [7, 8]. Furthermore, the electrical transport properties of the members of this solid solution are intriguing, in the sense that for $x > x_c$ we observe that the main conduction mechanism is variable range hopping (VRH) of carriers and the compounds exhibit GMR [8], while for $x < x_c$ the GMR effect vanishes rapidly with decreasing $x$ and for $x \leq 0.2$ the MR becomes positive [2]. As already mentioned no explanation has been suggested regarding the behavior of MR as a function of $x$ in the investigated solid solution.

There are several distinct differences between the perovskite manganites (R$_{1-x}$A$_x$MnO$_3$, R=rare earth, A=alkaline earth) and the LaNi$_{1-x}$Co$_x$O$_3$ solid solution: (i) The manganites are magnetically long range ordered systems [9] in contrast to the investigated system which exhibits ferromagnetic cluster glass behavior [8, 10]. (ii) The MR effect in manganites arises mainly from A-site substitution of the archetypal ABO$_3$ perovskite structure. It is widely accepted that substituting the Mn ion not only destroys the long-range ferromagnetic order but also suppresses the CMR effect [11]. On the other hand, the GMR effect in LaNi$_{1-x}$Co$_x$O$_3$ is due to B-site substitution. (iii) The maximum bulk MR in manganites, i.e. neglecting intergrain tunneling MR, occurs either at or just below the ferromagnetic ordering temperature [9], in contrast to the investigated system where the maximum MR appears well below the onset of the magnetic interactions [2, 8] and since it is not a magnetically ordered system the MR is believed to be a pure bulk effect. These apparent differences between the Co-doped nickelates and the extensively investigated manganites have prompted us to undertake the present study.

Recently, we have reported a comparative study between a semiconducting ($x$=0.8) and a metallic ($x$=0.4) member of the series [8]. It is clear that although the two compounds exhibit strikingly similar magnetic properties they display distinctly different electrical behavior. We have proposed a picture based on the existence of ferromagnetic Co based clusters and suggested that the cluster size growth induced by the applied magnetic field is responsible for the occurrence of the negative GMR effect for the semiconducting member ($x$=0.8); a fact which is consistent with previous studies on other cluster glass systems [12]. Another major conclusion of that study was the realization that an -$ln$T term is needed to fit well the resistivity data for the $x$=0.4 metallic compound.

The purpose of the present work is to investigate the low-temperature resistivity of several members of the LaNi$_{1-x}$Co$_x$O$_3$ solid solution with $x$ close to but below the critical concentration for the metal-to-insulator transition ($x$<0.65) in order to come up with a consistent picture for the magnetoresistive behavior of the metallic members of the series. Specifically, we investigate the temperature dependence of resistivity as a function of Co-doping and provide evidence that the suppression of spin fluctuations, to which we attribute the $ln$T dependence of resistivity, by the applied magnetic field, is the key mechanism in understanding the magneto-resistive behavior in the compounds with $x \leq 0.5$. Also, we show that the cluster percolation threshold (CPT) theory is relevant and apply it to analyze the resistivity data for the $x=0.6$ compound, which is just below the transition from the metallic to insulating state. The results of this analysis are consistent with the picture that the growth of the clusters in the presence of magnetic field is primarily responsible for the observed GMR effect in the $x=0.6$ compound.

## II. Experimental

Ceramic powder samples of LaNi$_{1-x}$Co$_x$O$_3$ ($x$=0.3, 0.4, 0.5, 0.6) were prepared by the Pechini (citrate-gel) method [13], using very high purity (99.999 %) metal nitrates [(La(NO$_3$)$_3$·6H$_2$O, Co(NO$_3$)$_2$·6H$_2$O, Ni(NO$_3$)$_2$·6H$_2$O] as starting materials. The nitrates were first dissolved in nano-pure water and then an appropriate mixture of citric acid and ethylene glycol was added to the solution. The gel formation was catalyzed by the addition of HNO$_3$. The resulted gel was decomposed at 400 °C and the acquired precursor powders were then calcined at 650 °C in a stream of oxygen gas for a few days followed by overnight annealing at 1000 °C in order to improve the crystallinity of the grains. The samples were subjected to a final heat treatment at 460 °C in ambient oxygen pressure. Iodometric titrations confirmed that the samples used in this study were optimally oxygenated (LaNi$_{1-x}$Co$_x$O$_{3\pm\delta}$, $\delta$=0.01).

The structure of the oxygenated powders was determined by x-ray powder diffraction using a Rigaku (RINT 2000) diffractometer with monochromated Cu $Ka_1$ radiation. Profile analysis and least squares numerical fitting of the obtained diffractograms that was performed in the hexagonal $R\bar{3}c$ space group showed no evidence for any impurity phases within the resolution of the instrument used. The extracted lattice parameters confirmed the formation of a solid solution. The cation stoichiometry of the samples was found to agree well with the nominal one using Energy Dispersive X-ray (EDX) spectroscopy.

Resistivity measurements were performed on bar-shaped dense pellets with the magnetic field applied in the plane of the contacts and perpendicular to the current direction. All data were collected in a homemade apparatus with a standard 4-probe dc technique in the temperature range $6 \leq T \leq 300$ K. As probes we used Au wires (ø = 30μm), which were attached to the samples with silver-paste (DUPONT, 4929). In order to reduce the resistance of the contacts we annealed the samples with the attached Au probes at 400 $^{o}$C for 2 hours in flowing oxygen. The I-V characteristics were linear at the lowest and highest temperature of measurement both in zero and nonzero applied magnetic fields.

### III. Results and Discussion

*A.) Compounds with x = 0.3, 0.4, and 0.5*

Figure 1 shows the temperature dependence of the zero-field resistivity, $\rho(T)$, for the $x$=0.3, 0.4 and 0.5 compounds. The resistivity of the $x$=0.3 compound (Fig. 1a) exhibits a linear-in-T behavior for T > 140 K and a minimum at $T_{min}$~45 K followed by an upturn at lower temperatures. Upon increasing the Co-doping to $x$=0.4, we observe a dramatic change in the behavior of the $\rho(T)$ curve (Fig. 1b). The location of the minimum in $\rho(T)$ is shifted to $T_{min}$~160 K and the slope of the high-temperature linear part is strongly decreased. Upon further doping ($x$=0.5) the $\rho(T)$ curve exhibits semiconducting-like behavior and no minimum is visible up to 300 K (Fig. 1c). Nevertheless, according to the authors of Ref. 5 the $x$=0.5 compound is metallic since the derivative $\partial ln\sigma/\partial lnT$ tends to zero at T=0.

The solid lines in Fig. 1 represent the fitting of the experimental $\rho(T)$ data to the following expression:

$$\rho(T) = \rho_0 + \alpha_1 T + \alpha_2 T^{1/2} + \alpha_3 lnT. \quad (1)$$

This expression contains four terms: the residual resistivity $\rho_0$ term [$\rho_0=\rho(T=0)$], which accounts for the scattering of conduction carriers by static defects such as grain boundaries; a linear term (~T) which accounts for scattering by phonons and is dominant at high temperatures; a $T^{1/2}$-term which arises from Coulomb interactions and is dominant at low temperatures; and an -$ln$T term which is characteristic of systems with spin fluctuations [14] and is observed over a limited range of low-temperatures below $T_{min}$.

The magnitude of $ρ_0$ (see Table I) obtained from the fitting procedure steadily diminishes with decreasing Co-doping and attains a value 1.12 mΩ·cm for $x$=0.3. This behavior is consistent with the fact that $ρ_0$~0.5 mΩ·cm for the metallic end member of the solid solution LaNiO$_3$ [3]. In addition, the magnitude of the coefficient $α_3$ decreases with decreasing Co-doping; this behavior can be attributed to the suppression of spin fluctuations with decreasing $x$ (*vide infra*).

In earlier studies the low-temperature conductivity $σ(T)$ of the metallic members of the investigated solid solution ($x ≤ 0.65$) has been found to exhibit an almost ~$T^{1/2}$ dependence and attributed to strong electronic correlations [5]. It is well known that in transforming the electron-electron (e-e) interactions conductivity formula to resistivity, higher order terms of the ~$T^{1/2}$ expansion should be used ($ρ_{e-e}(T)=ρ(0)-aT^{1/2}+bT-cT^{3/2}+…$) in cases where the measurements extend to high temperatures. However, the introduction of higher order terms in expression (1) did not result to any significant improvement of the fitting error value, which is a strong indication that electron-phonon interactions are mostly dominant at high temperatures. Thus Equation (1) should be viewed as the best fitting expression with the minimum number of physical parameters. For completeness, we mention that the coefficient, $α_2$, of the e-e interaction term in Equation (1), $ρ_{e-e}(T)$~$α_2T^{1/2}$, is equal to $α_2$=-$m_σρ_o^2$, where $m_σ$ is the coefficient of the e-e conductivity expression, $σ(T)$~$m_σT^{1/2}$, and contains the diffusion constant, $D$, and the screening constant for Coulomb interactions, $F_σ$ [15]. It is also worth mentioning that for all $x$ the fitting resulted in negative values for the coefficient $α_2$ in agreement with the transformed resistivity formula [15].

Figure 2 displays the resistivity-versus-$\ln$T curve for the $x$=0.3 compound measured in various magnetic fields. The existence of a linear regime clearly demonstrates that the resistivity exhibits an $\ln$T behavior over a significant range of low temperatures (the same holds true for $x$=0.4 and 0.5 and are shown as insets in Fig. 2).

The use of four parameters in the fitting formula of resistivity versus temperature data is usually not desired. However, there have been reports of systems with complicated scattering mechanisms where the usage of four terms was necessary to account for the experimentally observed curves. For example, Akimoto et al [16] have used the 4-parameter formula $\rho(T)=\rho_0+A_2T^2+A_3T^3+A_{9/2}T^{9/2}$ in order to speculate on the observation of single magnon scattering in chemically modified half metallic ferromagnetic manganites in the temperature range 5<T<100 K, and Gayathri et al [17] have used the formula $\rho(T)=\rho_0-\gamma T^{0.75}+\alpha T^{1.5}+\beta f(T)$ for LaNiO$_{2.86}$ in order to fit their data. The $f(T)$ term is a function of temperature to which the authors of Ref. 17 have attributed the form $\sim T^m$ with $m$=1 for T>100 K and $m$=2 for T<50K.

In our case a three terms formula ($\rho_0$, $\sim$T, $\sim T^{1/2}$), i.e. neglecting the logarithmic term, fits the data; however, such a fit results in an error between the experimental and calculated points that exceeds three to five times (depending on $x$) the statistical error of the measurement (<1%). It is noteworthy to point out that the authors of Ref. 5 were not successful in fitting their $\rho(T)$ data on LaNi$_{1-x}$Co$_x$O$_3$ samples with a single formula for the whole temperature range of measurements since $\sim T^{1/2}$ is effective at low temperatures (T<10 K) while the electron-phonon interaction term is significant at high temperatures as already discussed. Therefore, it seems that a physically justified, fourth term is required to reduce the mismatch. In our analysis we defined the quantity $\delta\rho = \rho_{exp} - \rho'_0 - \alpha'_1 T - \alpha'_2 T^{1/2}$, where $\rho_{exp}$ is the measured resistivity and $\rho'_0$, $\alpha'_1$, $\alpha'_2$ are the resulting coefficients from the three terms formula fitting (i.e. formula (1) without the $\ln$T term). We have observed a wide temperature-regime, ranging from 15 K up to 80 K depending on *x*, where $\delta\rho$ scales linearly with $\ln T$. In addition, the inclusion of the $\ln T$-term resulted in the acceptable fitting errors reported in Table I. Therefore we conclude that the $\ln T$-term is necessary in formula (1).

Moreover, we point out that the origin of magnetoresistance in the metallic members of the $LaNi_{1-x}Co_xO_3$ solid solution has never been discussed before. Using only the terms $\rho_0$, ~T and ~$T^{1/2}$ in the fitting procedure constrains us to attribute the observed MR to the e-e interaction term. However, Aleiner et al [18] have recently examined the problem of the quantum correction to the conductivity of strongly correlated systems and found that the resistivity minimum persists for all magnetic fields and its position is only weakly affected by them. In contrast, we observe a minimum that strongly depends on the magnetic field, i.e. increasing H results in a flattening of the minimum and a shift at lower temperatures. Therefore, we conclude that the observed MR is not an e-e interactions effect but rather stems from magnetic disorder, which is seen as an effective "spin-fluctuation" by the appearance of the logarithmic-in-temperature resistivity term.

Next we briefly discuss the possible origin of $\ln T$. The term "spin fluctuations" is quite general in its usage and may include several cases such as (i) the Kondo effect (ii) exchange enhanced systems (iii) localized spin fluctuations (LSF) and (iv) two band scattering [14]. The first two cases have been related to s-d exchange in **dilute** alloys of a non-magnetic host. However, the strong ferromagnetic signal observed in the investigated compositions as well as the unphysical values ($T_K<0$) of the Kondo temperature, $T_K$, deduced from fitting the incremental resistivity, $\Delta\rho(T) = \rho_{alloy}-\rho_{host}$, to the Appelbaum–Kondo relation [19], $\Delta\rho(T) \sim [(T/T_K)\ln(T/T_K)]^2$, suggest that the investigated system cannot be treated in the framework of the aforementioned models and thus cases (i) and (ii) will not be further discussed.

Cases (iii) and (iv) are two more typical examples of screening effects in s-d systems, that have been extensively discussed and applied in *concentrated* alloys [20]. Within the LSF model, the $\Delta\rho(T)$ is a logarithmic function of temperature according to the formula:

$$\Delta\rho(T) = A + B\ln[(T^2+\theta^2)^{1/2}], \qquad (2)$$

where $\theta$ is the spin fluctuations temperature [20]. We have performed such an analysis for the normalized incremental resistivity, $\Delta\rho'(T)$, of the $x=0.3$ compound:

$$\Delta\rho'(T) = \left.\frac{\rho(T)}{\rho(290K)}\right|_{x=0.3} - \left.\frac{\rho(T)}{\rho(290K)}\right|_{x=0} \qquad (3)$$

The $\Delta\rho'(T)$ has been used (instead of the incremental resistivity) in order to account for possible differences regarding the grain boundary resistances due to different packing densities of the pellets for the "alloy" and the "host". Figure 3 shows the temperature dependence of $\Delta\rho'(T)$ for $x=0.3$ in the temperature range $8 \leq T \leq 30$ K plotted as a function of $ln[(T^2+\theta^2)^{1/2}]$ in various applied magnetic fields. The best fit for H=0 T was obtained for $\theta=13.02\pm0.52$ K while this value drops to 9.76 K for $\rho(T, H=3T)$ and 8 K for $\rho(T, H=5T)$. Similar fitting to $\Delta\rho'(T; H=0)$ for x=0.4 gave $\theta \sim 11.5$ K.

The LSF model results from a modification of the conventional Kondo model assuming an additional relaxation of the "impurity spin", which in our case should be viewed as the magnetic moment of the ferromagnetic cluster, because of multiple scattering events via for example spin-orbit effects [21]. The physical meaning of $\theta$ is actually a measure of the inverse LSF lifetime, thus the decrease of $\theta$ with increasing magnetic field implies that the effect of the magnetic field amounts to the reduction of the fluctuation rates of the cluster moments by stabilizing them parallel to its direction.

Another plausible explanation stems from the relation $\rho \sim ln[(T^2+\Delta^2)^{1/2}]$, which describes the low temperature resistivity of amorphous alloys, i.e. structural two level systems [14]. In this case $\Delta$ is related to the energy difference between the two tunnelling states. Crystalline magnetic spin glasses are known to exhibit several physical properties at low temperatures (e.g thermal conduction), which are similar to those of amorphous materials [22]. It was thus proposed that the low-temperature behavior of spin glasses could be described in terms of magnetic two level systems, in an analogy to the structural two-level systems [22]. We suggest that a similar intra- or inter-cluster effect may take place in the $\pi^*$-$\sigma^*$ bands of the investigated system. If the latter is true, then $\varDelta$ should be related to a magnetic energy barrier, e.g. the magnetic anisotropy energy of the clusters, which gives rise to a random spatial variation of cluster moments that is in turn viewed as "spin fluctuations". It seems natural that decreasing the tunnelling barrier by the application of a magnetic field will result in lower resistivity values, and hence in negative magnetoresistance.

The above models can both explain the evolution of the slope of the solid lines, which fit the linear part of the $\rho$ vs. $lnT$ curves, as a function of the magnetic field in Fig. 2. It is evident that the higher the magnetic field the smaller the observed slope that pertains to suppression of spin disorder arising from spatial spin accumulation (clustering). It is important to mention that the negative MR values are higher in the metallic members of the solid solution for which spin fluctuations are stronger, i.e., they exhibit higher magnetization values and the coefficient $\alpha_3$ in Equation (1) is larger. We should emphasize that both models proposed here exhibit the same functional dependence, however, they are based on profoundly different physical backgrounds. An evident drawback of the second model is that the relation $\rho \sim ln[(T^2+\varDelta^2)^{1/2}]$ has been shown to be independent of the applied magnetic field for structurally amorphous alloys [23]. Therefore, one should consider it to be purely phenomenological. It is clear that further theoretical work is needed before one concludes on the validity of the second scenario although it is plausible.

To our knowledge, there have been thus far only three reports about oxides whose resistivity exhibits a logarithmic temperature dependence, which has been attributed to Kondo scattering [24-26]. In our opinion references [24] and [25], however, do not provide any conclusive data as to whether Kondo scattering is the origin of the logarithmic dependence of the resistivity. On the other hand, although the Kondo expression seems to fit quite well the data in reference [26], no accompanying measurements, like thermoelectric power, are reported which could elucidate the nature of the Kondo interaction in this oxide. It is important to note that regardless of which of the above cases may be responsible for the *ln*T dependence, all of the theoretical models have been developed for metallic s-d exchange systems, i.e. wide band metallic hosts, which are in marked contrast to 3d or 4d-O 2p hybridized oxide metals. Therefore, we believe that this physical effect deserves further theoretical as well as experimental investigations.

Finally, we should briefly refer to the anomalous behavior of the fitted parameters observed in Table I, i.e. $\alpha_2$ is equal for *x*=0.3 and 0.5 while it increases for *x*=0.4. Although, this may be considered as an artifact of the fit we refer the reader to a similar "anomaly" regarding the *x*=0.4 compound in Table III of Ref. 5. The authors of Ref. 5 have used the e-e interactions formula $\sigma(T)=\sigma_0+\alpha T^m$ to fit conductivity data for T<2 K and 0<*x*<65. They have found that although ***α is essentially the same for compositions x=0.4 and 0.57*** the exponent *m* differs quite significantly, i.e. ***m=0.31 for x=0.4 and m=0.53 for x=0.57***. Therefore, we conclude that this "anomaly" observed for the x=0.4 compound stems from the underlying physics of the compound rather than being an artifact of the fit. On the other hand, the anomalously large $\alpha_3$ coefficient for *x*=0.5 is likely related to the low temperature magnetic state of the specific compound which most probably is on the verge of becoming a long-range order ferromagnet [27, 28]. Therefore, we expect spin disorder scattering to be quite increased in such case.

B.) *Compound x=0.6*

Figure 4 displays the temperature dependence of the resistivity for the x=0.6 compound. This member of the series is a case of special interest because even though its $\rho(T)$ curve exhibits semiconducting-like behavior and the resistivity values are enhanced almost 2 orders of magnitude in comparison to *x*=0.5 compound, it is considered to be metallic according to the authors of Ref. 5. In addition, it has been claimed that for this doping level (regime near the metal-to-insulator transition) there is no theory that can explain the data [2, 5]. We show next that there is enough strong evidence, which can justify the application of the CPT theory.

CPT theory has been used over the years to describe a wide variety of phenomena, including conductivity of random resistor networks, gelation of polymers and smoke particle aggregation [29]. A new application of CPT theory, with significant impact is the explanation of metal-to-insulator transitions in transition metal oxides [30]. We claim that the CPT theory is applicable in the case of the $LaNi_{1-x}Co_xO_3$ system based on the following facts: (i) CPT theory applies to solid solutions and requires that one end member of the solution is metallic and the other is insulating, and that the two phases interpenetrate each other randomly. This is indeed the case for the investigated solution as we have mentioned in the introduction and experimental parts. In addition, we have shown that a cluster glass phase exists throughout the whole compositional range. These clusters seem to remain fairly isolated at low Ni concentrations, while they exhibit a substantial overlap at *x*<0.4 [8]. (iii) CPT theory predicts, for a cubic lattice, that a metal-to-insulator transition occurs at a critical concentration $x_c$=0.69 (or $x_c$=0.31 for transition from insulating to metallic state) [31]. Note that for the investigated system $x_c$=0.65. The small deviation in the value of $x_c$ is attributed to the non-cubic structure of the members of the investigated solid solution. (iv) CPT theory predicts variable range hopping of charge carriers in the insulating regime above the percolation limit. Our resistivity data on $LaNi_{0.2}Co_{0.8}O_3$ [8] and $LaNi_{0.3}Co_{0.7}O_3$ (unpublished) provide concrete experimental evidence for the occurrence of VRH for $x>x_c$. (v) Kirkpatrick has shown that the conductance in truly percolative systems scales as $G(x) \sim (x-x_c)^n$ with $1.5 \leq n \leq 1.6$ [32]. In fact, because the above relation was extracted for cubic systems, $n$ rarely takes the above-predicted values and instead varies between 1.5 and 2.1 [32, 33]. A typical example is the case of $Na_xWO_3$ for which $n=1.8$ [32]. Regarding the investigated system, we have found that the zero point conductivity, $\sigma_0$, scales as $(x-x_c)^2$ (see inset of Fig. 4). We conclude that the aforementioned arguments (i)-(v) constitute concrete evidence to classify the metal-to-insulator transition in the investigated solid solution as a percolative phenomenon, i.e. a purely geometrical effect. Finally, we comment that in order to conclude on an accurate exponent describing the percolation procedure in the investigated solution would require several points of $x$ close to $x_c$, however, such a study lies beyond the scope of the present paper.

From the above it becomes evident that compounds with $x$ just below $x_c$ can be considered as a blend of two components, a metallic and an insulating one, with conductivities $\sigma_1$ and $\sigma_2$, respectively. In such a case the overall conductivity $\sigma_{tot}$ (i.e., the measured conductivity), is a function of the relative volume fraction of the two phases as well as the shape and distribution of the cluster particles or domains of each phase [34]. Since the mean free path, $\ell$, is sufficiently small compared to the relative size of the particles [35], one can assume that the cluster particles are immersed in a homogeneous medium of conductivity $\sigma'$ and use the following expression, which holds for spherical clusters [36, 37]:

$$V_1 \frac{\sigma' - \sigma_1}{\sigma_1 + 2\sigma'} + V_2 \frac{\sigma' - \sigma_2}{\sigma_2 + 2\sigma'} = 0. \qquad (4)$$

$V_1$ and $V_2$ are the volume fractions of the metallic and the insulating phases, respectively ($V_1=1-V_2$). In order to proceed with the calculation of the relative volume fraction of the two phases we consider $\sigma'=\sigma_{tot}$, i.e., we assume that the particles are embedded within a homogeneous effective medium of conductivity $\sigma_{tot}$ which will be determined self-consistently. Another possibility would be to consider the case $\sigma'=\sigma_2$, but this is unsuitable since the relative concentration of the insulating phase (LaCoO$_3$) is comparable to that of the metallic phase. The calculation of $\sigma_1= 1/\rho_1$ was based on a high-temperature fit (T>100 K) of the resistivity data shown in Fig. 4 to the following expression, which represents the metallic phase:

$$\rho_1=0.055 + 3\times10^{-5} T - 0.011\, lnT. \qquad (5)$$

On the other hand, $\sigma_2=1/\rho_2$ was obtained by fitting the resistivity data in the low temperature regime (T<40 K) to the VRH expression [8], which represents the insulating phase:

$$\rho_2=38.855 \times exp\{-2.772/T^{1/4}\}. \qquad (6)$$

In both cases the maximum fitting error does not exceed 1.1%. Replacing in equation (4): $\sigma'=\sigma_{tot}$, $V_2=1-V_1$ and $\sigma_1=1/\rho_1$, $\sigma_2=1/\rho_2$ with expressions (5) and (6), respectively, we obtained the volume fraction of the metallic phase, $V_1$, as a function of temperature (shown in Fig. 5). Two important observations can be made: first that the low temperature value of the metallic volume roughly corresponds to the concentration of Ni cations in the matrix and second that $V_1$ grows to almost 100% at ~100K. As evident from Fig. 4 for T>100K the resistivity for $x=0.6$ (nominal [Ni$^{3+}$]=0.4) drops two orders of magnitude and thus the compound can be considered purely metallic. It is noteworthy to point out that a similar analysis of resistivity data taken in the presence of a magnetic field of 45 kOe shows that $V_1$ is enhanced by 3% relative to zero field values up to 80K. Evidently, the increase of $V_1$ in the presence of the magnetic field lends further support to the proposed picture that spatial growth of magnetic clusters is primarily responsible for the observed phenomenon of GMR in $x \geq 0.6$ compounds [8].

In order to further check the validity of the above results, we considered a naive application of Matthiessen's rule in the sense that the random interpenetration of the aforementioned (metallic + non-metallic) phases can be expressed using the expression for a series connection of resistances:

$$\rho(T) = \rho_{metal} + \rho_{semiconductor}, \qquad (7)$$

where $\rho_{metal}$ corresponds to equation (5) and $\rho_{semiconductor}$ to equation (6). Subsequently, we used the following equation:

$$\rho(T) = \rho_0 + \beta_1 T + \beta_2 \ln T + \beta_3 \exp(-T_0/T)^{1/4}. \qquad (8)$$

The solid line in Fig. 3 is the result of fitting the data of x=0.6 compound using equation (6). The maximum fitting error of the curve is 2.8% while the calculated goodness of the fit is $R^2=0.9990$. We should mention that compositions close to x=0.65 have been previously reported to obey for a limited temperature range the empirical equation $\rho(T) = \rho_0 \exp(E_a/k_B T + \theta)$ which, nevertheless, is physically unjustifiable [38].

## IV. Conclusions

In conclusion, we have undertaken a detailed study of the resistivity of the LaNi$_{1-x}$Co$_x$O$_3$ solid solution for $0.3 \leq x \leq 0.6$ in the temperature range 6-300 K. For the compounds with $0.3 \leq x \leq 0.5$ we found the existence of a $-\ln T$-dependence of the resistivity which has been attributed to the existence of spin fluctuations. In addition, we interpreted the decreasing contribution of the $-\ln T$ term to resistivity with increasing magnetic field as the suppression of spin fluctuations by the magnetic field and therefore, concluded that this mechanism is primarily responsible for the magnetoresistive behavior of these compounds. The existence of a logarithmic term in the resistivity expression of an oxide is mostly unusual as well as intriguing and it's warrant of further investigations.

Furthermore, we provided concrete evidence that the metal-to-insulator transition in this solid solution is a percolation phenomenon. We applied the CPT theory to analyze the resistivity data for the $x=0.6$ member and found that the metallic volume of the matrix grows to 100% around 100 K while it decreases to almost the nominal [Ni] concentration (40%) at low temperatures. Also, we found that the metallic volume in the presence of a 45 kOe field increases by ~3% compared to the zero field value.

**FIGURE CAPTIONS**

**Figure 1.** Temperature dependence of the resistivity in zero magnetic field for (a) $x$=0.3, (b) $x$=0.4 and (c) $x$=0.5. The solid lines are fittings to the data using equation (1). In order to make the fitting clearer we have removed some of the experimental data points from the figure.

**Figure 2.** Resistivity versus $lnT$ plots for $x$=0.3 at different applied magnetic fields (indicated on the graph). The linear fittings provide concrete evidence for the $lnT$ dependence. Notice that the higher the field the smaller the slope of the fitted line.

**Figure 3** Inset shows the plot of normalized incremental resistivity $\Delta\rho'$(T) vs $ln[T^2+\theta^2]^{1/2}$ for x=0.3.

**Figure 4.** Temperature dependence of the resistivity of the x=0.6 compound. The solid line is a fit to the data using equation (6) according to which the measured resistivity can be considered as a blend of a metallic and a semiconducting component. The inset shows that the zero point conductivity of the investigated solid solution scales as $(x-x_c)^2$. Filled circles are our data while open squares were taken from Table III of Ref. 5.

**Figure 5.** Temperature dependence of the metallic volume fraction, $V_1$, for the x=0.6 compound. It is noteworthy to mention that $V_1$ grows to almost 100% at high temperatures (T>100K), while its value at the lowest measured temperature (T=6 K) is slightly lower than the nominal concentration of Ni cations (40%).

Table I. Fitting parameters of the resistivity data in the temperature range 6≤T≤300 K for the LaNi$_{1-x}$Co$_x$O$_3$ compounds with $x$=0.3, 0.4 and 0.5.

|  | Composition | | |
| --- | --- | --- | --- |
| *Parameter* | *x=0.3* | *x=0.4* | *x=0.5* |
| $\rho_0$ *(mΩ.cm)* | 1.1 ± 0.0 | 2.1 ± 0.0 | 9.3 ± 0.0 |
| $\alpha_1$ *(x 10$^{-6}$)* | 2.4± 0.0 | 2.8 ± 0.0 | 7.8 ± 0.0 |
| $\alpha_2$ *(x 10$^{-5}$)* | -3.01 ± 0.04 | -8.0 ± 0.1 | -3 ± 1 |
| $\alpha_3$ *(x 10$^{-6}$)* | -6.5±0.8 | -9.8±0.3 | -1690±3 |
|  | *Characteristics of the fit*[*] | | |
| N | 978 | 327 | 856 |
| $\chi^2$ | 1.74x10$^{-12}$ | 1.13x10$^{-11}$ | 1.75x10$^{-9}$ |
| $R^2$ | 0.9998 | 0.9985 | 0.9983 |
| *Error (%)* | 0.3 | 0.24 | 2.5 |

[*]$N$ is the total number of experimental data points, $\chi^2=(1/N)\Sigma(\rho_{obs}-\rho_{fit})^2/\rho_{fit}^2$, $R^2$ defines the goodness of the fit, $Error(\%) = 100 \times (\rho_{obs}-\rho_{fit})/\rho_{obs}$

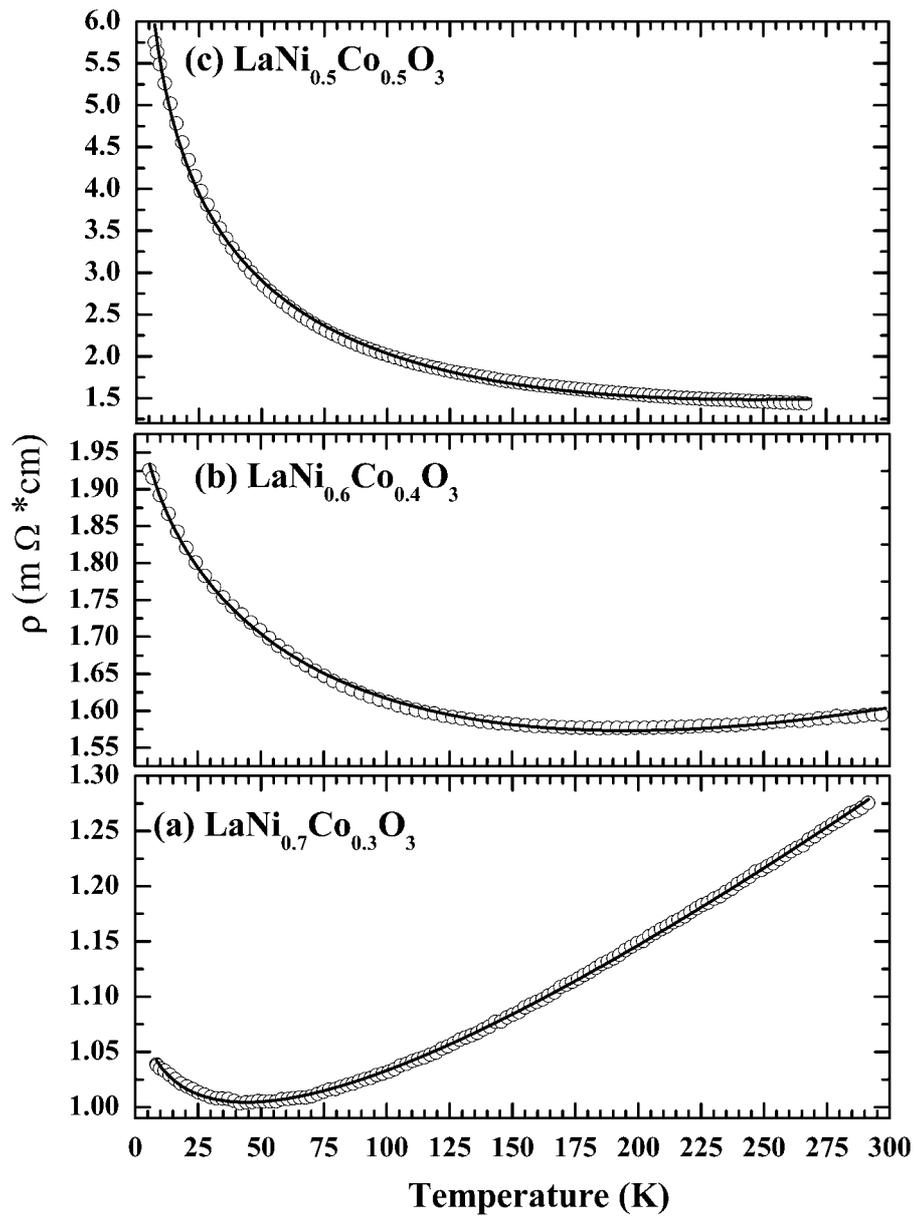

FIGURE 1: J. Androulakis et al, J. Magn. Magn. Mater.

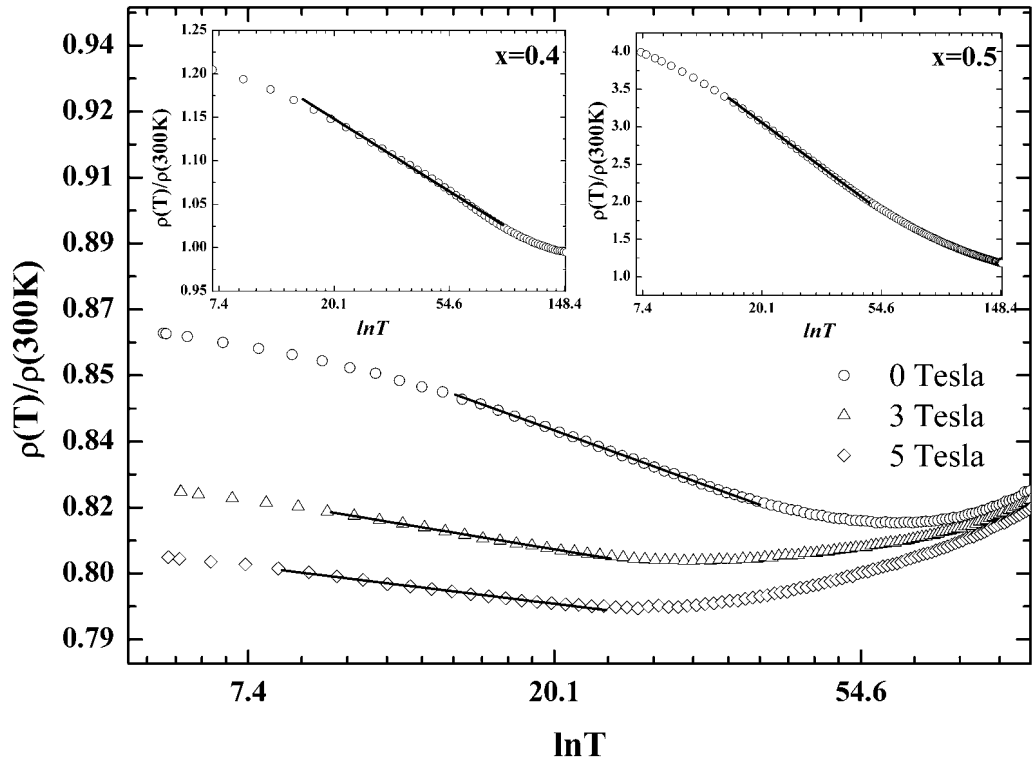

FIGURE 2: J. Androulakis et al, J. Magn. Magn. Mater.

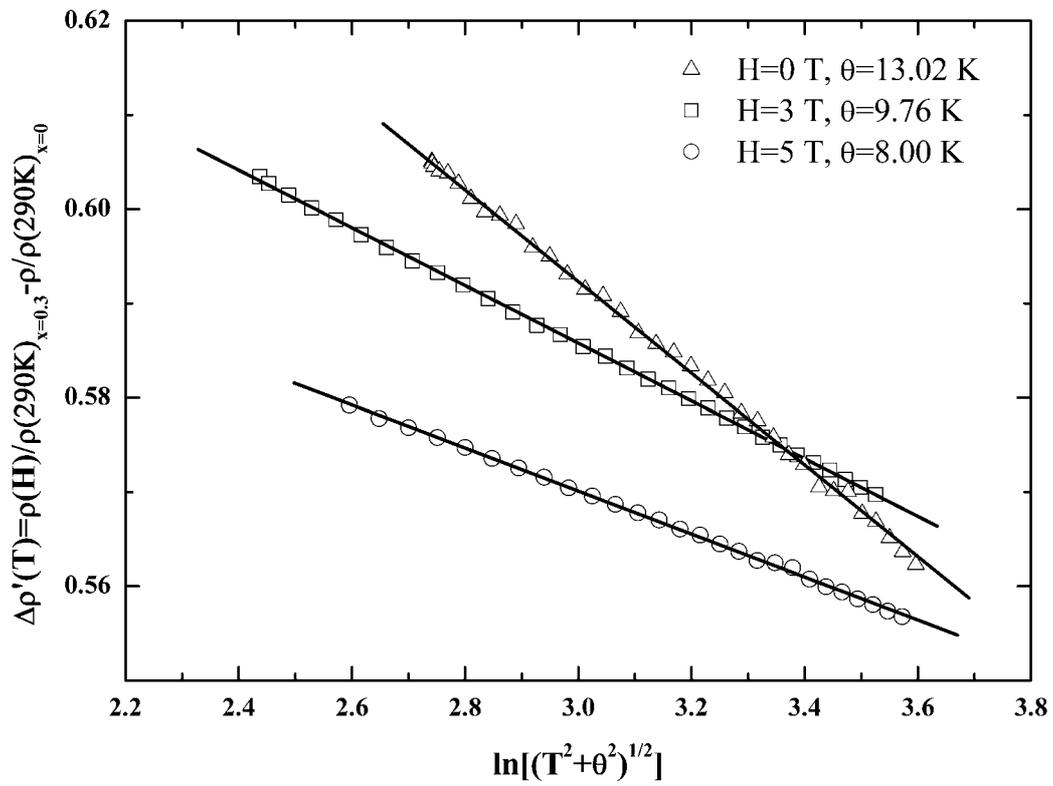

FIGURE 3: J. Androulakis et al, J. Magn. Magn. Mater.

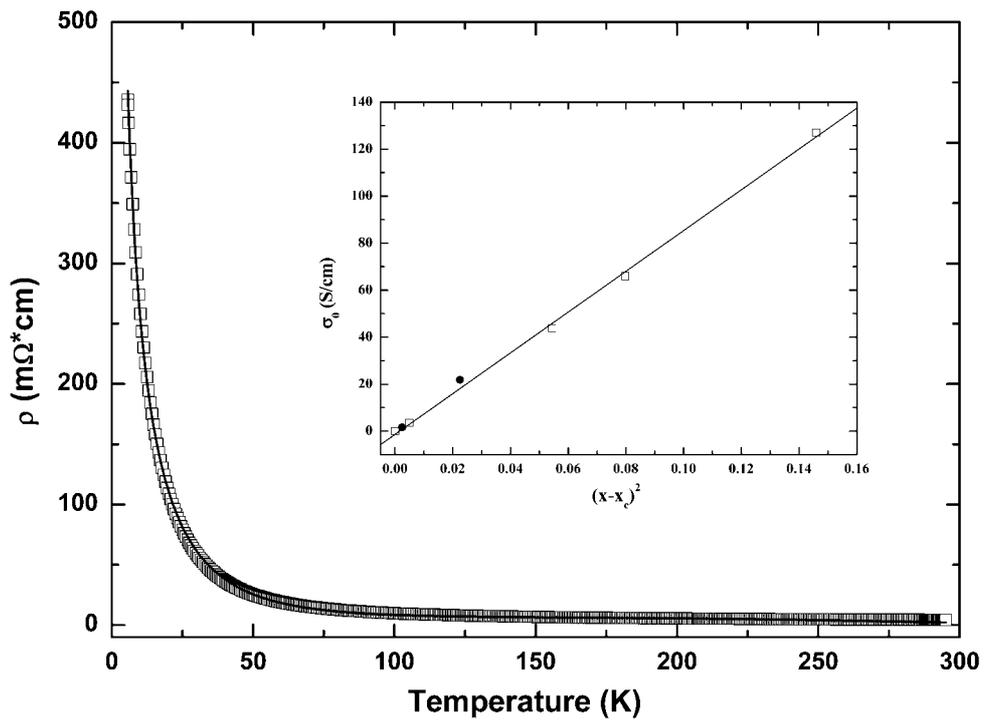

FIGURE 4: J. Androulakis et al, J. Magn. Magn. Mater.

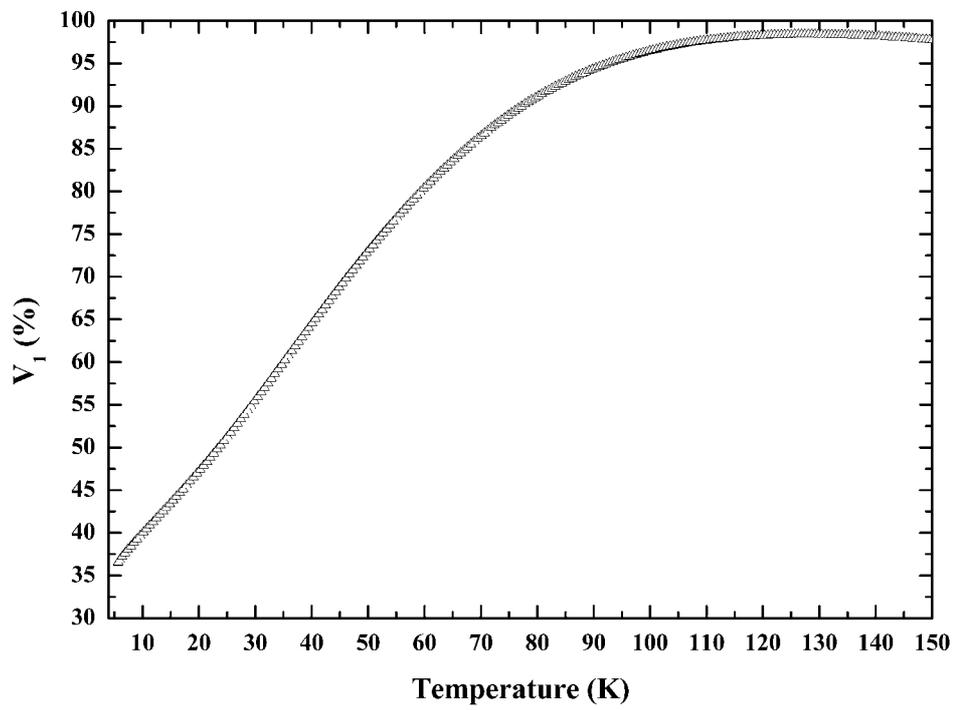

FIGURE 5: J. Androulakis et al, J. Magn. Magn. Mater.